# A Comparative Analysis of Supportive Navigation on Movie Recommenders

*Mohammad Sualeh Ali, Muhammed Maaz Tariq, Alina Ahmed, Abdul Razaque Soomro, Danysh Syed*

## Abstract

This literature review covers the research and thought process that went into making a solution for the infinite scrolling problem faced in streaming services such as Netflix. Using the data collected, we have come to the conclusion that an alternate layout can somewhat alleviate the problems it takes in navigating a list of movies. We also found out by a comparative analysis that some layouts, the circular one in particular, is advantageous in certain settings making it an ideal candidate for a movie recommender system.

## Problem Statement

The problem faced here is the Infinite Scrolling Problem, in which users of Netflix failed to find the show that they were looking for and instead spent way more time scrolling than actually watching something. And the few that managed to find a movie usually found it because of their pre-existing knowledge and biases and the scrolling activity did little to help them. This often turned into long mundane periods of tireless scrolling activity, with nothing of value being added to their movie knowledge-base in between.

Ultimately, what would happen was that the user having invested a lot of time in the scrolling activity could not commit to any one particular movie and would decide that the two hour investment was not worth it and would retire from the activity altogether ultimately becoming too overwhelmed and indecisive about the choices presented to them.

The fact that users find a platform relevant for their use is what stimulates revenue generation for the providers of that platform. The more agitated the users feel when they aren't able to find the content they need on a particular platform after a lot of scrolling, or maybe finding it, but after spending too much time on scrolling, the less likely they are to visit that platform in the future. This could result in substantial losses for these online businesses who thrive on user presence on their respective platforms - both directly and indirectly. Losing their customers would mean people would unsubscribe from their services, which would be a direct loss to their business, and furthermore, would act as a deterrent for the advertisers to put their ads on their platform. In short, just the fact that users have to scroll a lot in vain, could be very harmful to content providing business like Netflix. Here's where our solution comes in: a Circular Layout. A circular layout is rather like a reel of film that unfolds in a circular fashion: instead of moving vertically or horizontally to see more content (movies in our case), we'd be rotating a wheel-like view to browse through content. This view, as opposed to a grid-view for example, does not require screen-length swipes which are analogous to long paper scrolls. Rather, the user moves the wheel-view in a circular motion of the finger analogous to what we'd do to unroll a reel of paper. This layout is therefore, more compact and the users would be less likely to feel they have browsed a lot. Moreover, the way we have designed our circular layout, allows the user to view the details of the movie without making the screen look cluttered. With more details on the page, organised in a cleaner fashion, we have hypothesized that the users would hit the desired movie in a lesser amount of time, without being overwhelmed by the amount of content they've viewed.

## Literature Review

The current problem is that the grid interface of Netflix spoils the user in terms of choice [5]. The user is lost in the infinite scrolling and more or less intimidated by the never ending wall of movies and tv shows. We found that infinite scrolling can be downright harmful to usability — in particular, for search results and on mobile. However, it's not black and white, because the performance of each method varies according to the context of the page [1]. Content loading is also more efficient with infinite scroll. When users read contents displayed on the screen, content below are being loaded in the background. Once users scroll down or press "page down", contents will be displayed immediately [2].

The realization is well felt that having endless amounts of content to watch brings with it a constant feeling, that there is always something better waiting to be viewed [5]. However, initial results received relatively less exposure. Infinite scrolling is, therefore, ideal for quickly showing the breadth of an entire category; but because users aren't naturally halted when scrolling, they tend to scan more and focus less on individual products on the list [2]. In the end, one may just stick with what they know, play it safe. Or unable to make a decision at all and do other things [5]. Pinterest-style infinite scroll layouts design will keep on displaying content when users scroll down. Users will not reach an "end" of a page and there is no need to click "next page" for new content. Turning for the next page can be a distraction to the browsing experience, because users need to explicitly click a button or an arrow, and then wait for the next page to load. [1]

Users' contributions: recommending products, posting reviews, reading other users' contribution, interacting with other users, implementing systematic feedback systems and providing virtual co-presence features [1].

Suggestions in whatever form, overall purpose is to divert and hold our attention long enough for our interest to catch up [5]. Since various observations show that humans' ability to perform multiple tasks at the same time is limited, the capacity theory of attention assumes that the total amount of attention which can be used at a specific time is limited. Parallel performing multiple tasks is possible (e.g., driving and talking), while the attention capacity is divided among the activities. When two activities demand more attention capacity than is available, only one activity can be completed [2].

What the author suggests in [5] is 'This menu will display a clip of characters and actors from different movies and shows in one fixed scene depending on genre creators. In this scene, the crossover of characters will be displaying a sense of personality and characteristics from their particular movie or TV show.' This, despite being a good idea, is unrealistic because it would cost a great amount of money and graphic designers to make such a scene come to life. The copyrights to display those characters in their environment would prove to be another hurdle for Netflix.

The Pinterest-style infinite scroll layouts have two major components: dynamic grid layouts and infinite scrolling [5]. The main characteristic of dynamic grid layouts is that each grid has dynamic size in width and height, and all grids are arranged to fill in every blank space available. Dynamic grid layouts crams as much content as possible onto a single page, which is presumably justified by a statement like, "the more we show users, the more likely they are to find something they like." [2]. Infinite scrolling, also known as "endless scrolling", is a technique where additional content for a web page is appended dynamically to the bottom of the page as the user approaches the end. Users are less likely to continue on to the next "page" if they have to click something versus it being delivered automatically to them. Infinite scrolling will increase the average time spent on websites by users [2].

## Process Flow

For the process flow we first decided to conduct extensive research on all of the available material of scrolling layouts and movie recommender systems in general. After that we decided to create a tentative layout of a circular model which was then put into a pilot run. To evaluate its feedback we circulated the app amongst a representative sample to test it out.

One of the critiques we found out was that a lack of a comparative baseline to gauge its effectiveness. To make sure the yardstick was consistent across all the layouts we decided to portray the same set of movies across the three layouts: circular, grid, and a sliding layout. Based on this, we calculated both objective and subjective metrics to determine the sentiments of the user and which did they prefer as per their User Experience. This involved taking a post-survey after being shown three layouts and also calculating metrics based on their touch interaction with the interface.

A diagrammatic overview of our process flow is given below:

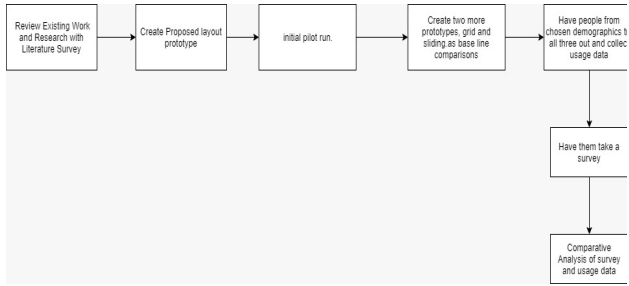

Figure 1: Process Model of the Research

## Proposed Model

### Audience

- Young adults
- Senior citizens
- Uneducated / Illiterate people

### Model

#### Circular Layout

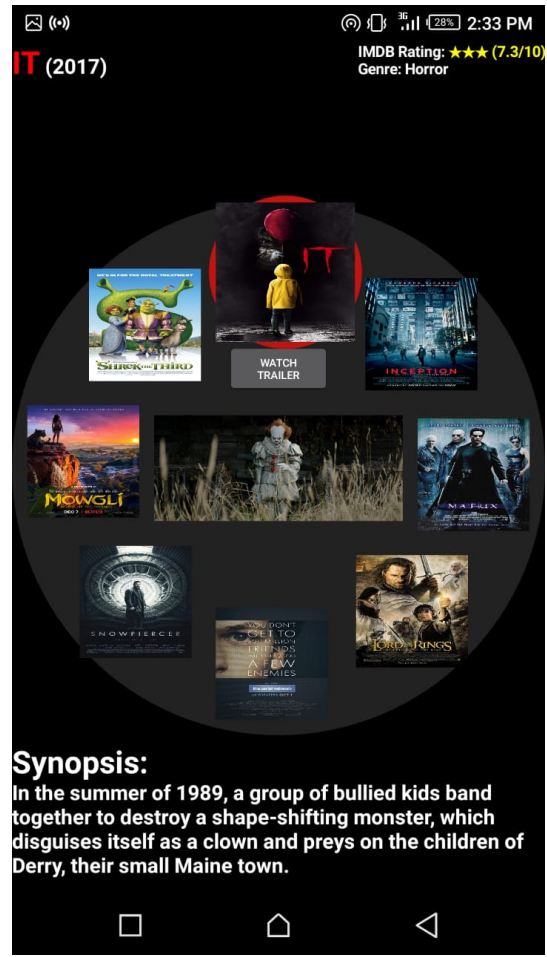

Figure 2: A screenshot of circular layout

This model is implemented using a rotating wheel that upon rotation clockwise procedurally loads new recommended movies unto the wheel. Similarly, the converse action i.e rotating it anti-clockwise can allow you to cycle the movie backwards. At any given time, there is a predefined limit as to how many movies can be loaded onto the wheel. For the purpose of symmetry we have currently set it to eight but this is an arbitrary number and can be changed depending upon factors like screen size, size of thumbnail , and so on.

As you can see from the figure above, we have intentionally created a donut-like interface in the centre for loading additional information like rating score, genre of movie, a brief synopsis of the movie, and a slideshow of scenes from the movie so that the scrolling motion can facilitate additional content. Furthermore, scrolling through the movies gives a brief glimpse of each movie's details, which may lead to an interest in movies that may otherwise not be paid attention to. Normally, you are only able to see the thumbnail and the movie title during the scrolling motion, while the circular layout adds additional information as well.

Additionally, the recommended movies are all at an equal distance from the center, providing equal importance to each movie in terms of their position in the layout. According to , a statistically significant improvement in the number of clicks was observed over grid and list when testing the circular layout.

Grid

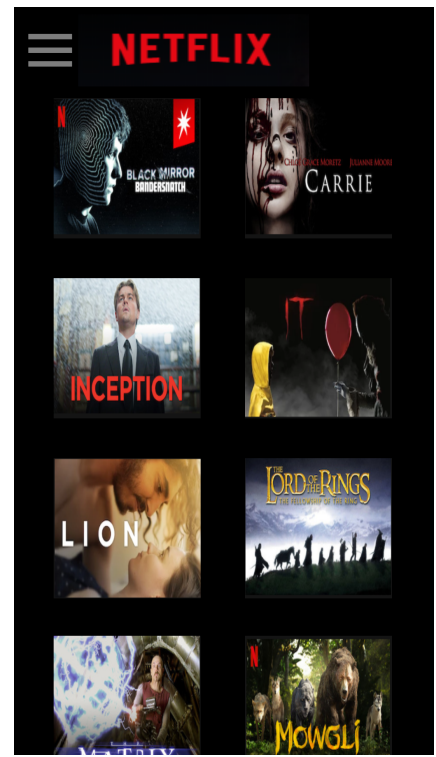

Figure 3: A screenshot of grid layout

We used a grid layout as a baseline because it is the preferred layout for most movie recommender systems. The movies are laid out in an N-dimensional grid and scrolling involves moving the grid progressively back and forth until you find the preferred movie.

Sliding

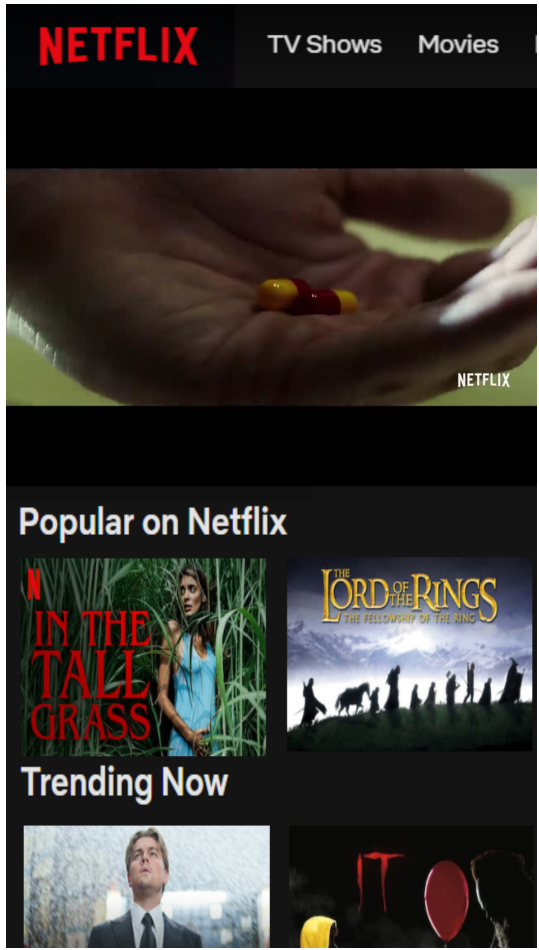

Figure 4: A screenshot of swim-lane/sliding layout

Similarly, our second baseline was the sliding menu in which the movies are listed in a sequential manner either horizontally or vertically. In this case we opted for horizontal swim-lanes to mimic a Netflix-based layout.

# Experiment Setup

Objective Measures

- First click
- Last click
- No. of movies shown interest in
- No. of clicks

Subjective Measures

| Variables | To be measured comparatively for the three choices |
|---|---|
| Decision confidence | Q: I am confident that this interface is the best choice. 1)Grid 2)Pie 3)Sliding Q: I am confident this interface is my least favourite choice. 1)Grid 2)Pie 3)Sliding |
| Perceived recommender interface's competence | Q: I found it easy to use this interface to search for movies. 1)Grid 2)Pie 3)Sliding Q: This interface is easy to maneuver. 1)Grid 2)Pie 3)Sliding Q: This interface provides better mobility. 1)Grid 2)Pie 3)Sliding |
| Enjoyability | Q: This interface was relaxing to my eyes. 1)Grid 2)Pie 3)Sliding Q: This interface was not frustrating to use. 1)Grid 2)Pie 3)Sliding Q: This interface is too cluttered. |

|  | 1)Grid 2)Pie 3)Sliding<br>Q: This interface helped me find good movies.<br>1)Grid 2)Pie 3)Sliding |
|---|---|
| **Behavioral Intention** | Q: I am inclined to use this interface again.<br>1)Grid 2)Pie 3)Sliding<br>Q: I would choose this interface for other apps if it were an option.<br>1)Grid 2)Pie 3)Sliding |

# Survey Results

## Preliminary Analysis

We did an initial pilot run with the circular layout and and took a survey of 17 respondents with their responses recorded in. The following chart shows the frequency distribution of our target audience:

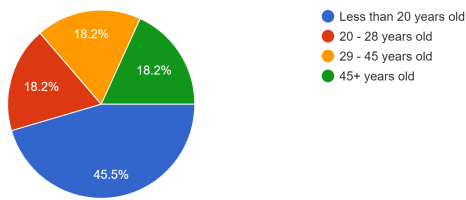

Figure 5: Age breakdown of surveys

Therefore, most of our respondents were young adults - the category of populace that is the major user of apps like Netflix.

As you can see in the figure below, a significant majority (76%) thought that the inclusion of additional information was helpful in making their decision. And overall, 71% of our respondents thought that the circular layout was a good approach in finding movies.

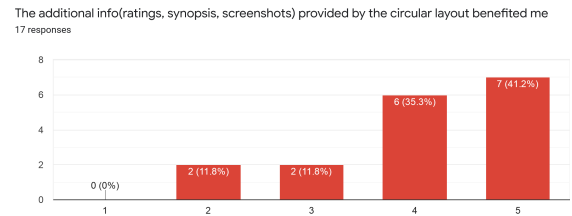

Figure 6: Survey response on helpful information provided by the Circular Layout

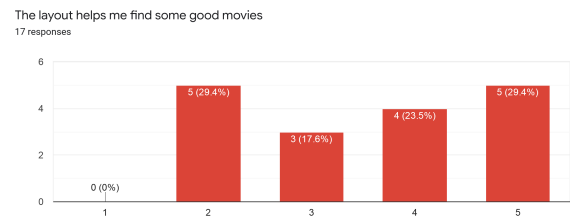

Figure 7: Survey response on assistance provided when searching for good movies.

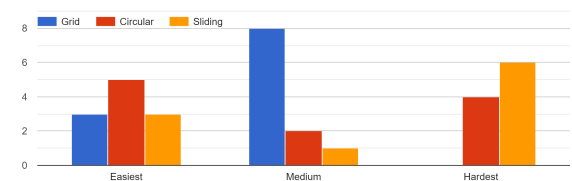

Figure 8: Survey response on ease of use when searching for movies

The survey response above shows that most people found the circular layout easiest to use, while most people found the grid interface to be moderately easy to use. Sliding interface was considered to be the least easy and the most difficult layout to use.

## Secondary Analysis

We can see from the results that circular layout benefits from Fitts Law [3] because it enables faster response time as distance to the movies from the centre is minimized and large size of the movie thumbnails allows more space for touch interaction.

### Analyses of Objective Metrics

The objective metrics, which we collected through the users' interaction with our three layouts - circular, grid, and sliding - seem to support our hypothesis that circular layout is better than the other two. For example, consider the following bar chart:

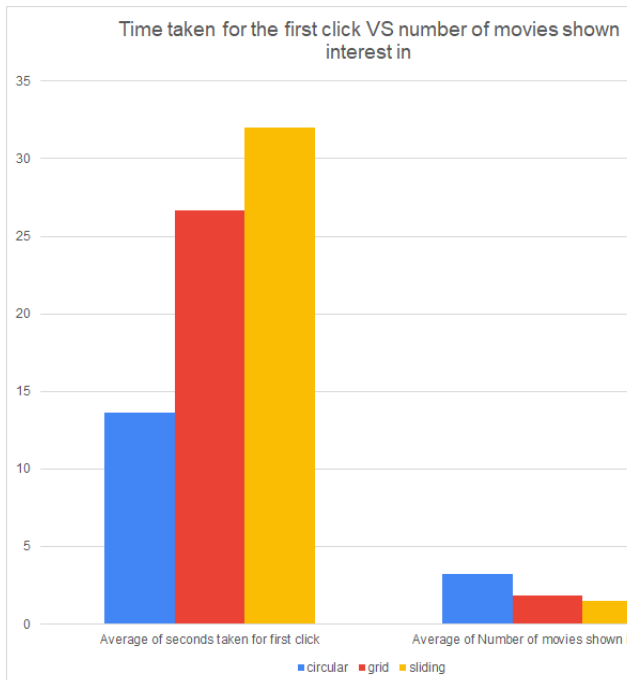

Figure 9: A comparison of average time taken amongst layouts

The graph shows two bar charts: one for average number of movies the user has shown interest in using a particular prototype, and the other for the average number of seconds taken by the user to click after starting the app for each of the 3 prototypes. The bar chart shows that on average, a less interested user takes more time to click on their first movie. There is apparently an inverse relationship between the two averages plotted on the graph above. More particularly, the users showed interest in more movies when they navigated using a circular layout, and hence, while using this layout, they were more likely to click earlier on some movie. This goes hugely in favour of our hypothesis, as it proves to some extent that circular layouts lead to fruitful user engagement with the app.

The diagram on the next page gives another interesting insight. It shows that the average number of seconds between the first and last click of the user varies *significantly* with the layout they are using. This metric measures that time taken by the user to find/decide what movie to watch. The bar chart shows that while using a circular layout, the user would, on average, take around 18x less time than grid layout, and 10x less time than sliding layout to find/decide the movie to watch. This means that users do not waste a lot of time mindlessly scrolling through the app while using a circular layout.

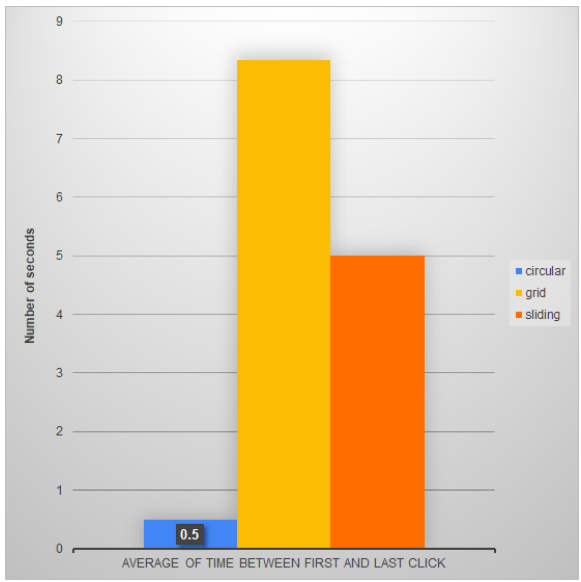

Figure 10: Time Comparison of First And Last Click with different layouts

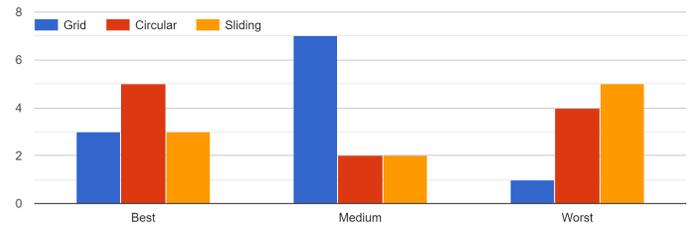

Figure 11: Survey response on best mobility provided by each interface, ranked from best to worst.

# Future Work

One of the surprising results that our survey summary showed was the polarization in users' feeling about the mobility of circular interface. This could be due to unfamiliarity with the interface causing some people to have a harder time with the interface compared to others. Another factor could be lack of respondents,. We had a limited number of survey respondents and it is possible that testing on a larger scale could give us more information. A future study could further study these trends and the reasons behind this polarization.